# Charge ordered ferromagnetic phase in $La_{0.5}Ca_{0.5}MnO_3$.

James C. Loudon, Neil D. Mathur & Paul A. Midgley

*Department of Materials Science and Metallurgy, University of Cambridge, Pembroke Street, Cambridge CB2 3QZ, UK*

**Mixed valent manganites are noted for their unusual magnetic, electronic and structural phase transitions. The $La_{1-x}Ca_xMnO_3$ phase diagram[1] shows that below transition temperatures in the range 100–260 K, compounds with $0.2 < x < 0.5$ are ferromagnetic and metallic whereas those with $0.5 < x < 0.9$ are antiferromagnetic and charge ordered. In a narrow region around $x = 0.5$, these totally dissimilar ground states are thought to coexist[2,3]. Uehara et al.[4] have shown that charge order and charge disorder can coexist in the related compound, $La_{0.25}Pr_{0.375}Ca_{0.375}MnO_3$. Here, we present electron microscopy data for $La_{0.5}Ca_{0.5}MnO_3$ that sheds light on the distribution of coexisting phases and uncovers a novel and unexpected phase. Using electron holography and Fresnel imaging, we find micron sized ferromagnetic regions spanning several grains coexisting with similar sized regions with no local magnetisation. Holography shows that the ferromagnetic regions have a local magnetisation of $3.4 \pm 0.2$ $\mu_B$/Mn (the spin aligned value is 3.5 $\mu_B$/Mn). We use electron diffraction and dark field imaging to show that charge order exists in regions with no net magnetisation and, surprisingly, can also occur in ferromagnetic regions.**

Fig. 1 shows the temperature dependence of the magnetisation for three different compositions of polycrystalline $La_{1-x}Ca_xMnO_3$ with a grain size of 5 μm produced commercially by Praxair. These measurements were made in a field of 1 T with a vibrating sample magnetometer. The $x = 0.3$ compound (Fig. 1(a)) shows a standard paramagnetic to ferromagnetic phase transition at 256 K and approaches the theoretical saturation moment of 3.7 $\mu_B$/Mn (derived from the filling of the Mn d levels) at low temperatures (down to 10 K) as expected. The magnetisation of the $x = 0.66$ compound (Fig. 1(c)) shows a paramagnetic to antiferromagnetic phase transition at 250K. The $x = 0.5$ sample from which the electron microscopy data is taken appears to show both a paramagnetic to ferromagnetic transition at 233 K and then an antiferromagnetic transition at 120 K measured on the cooling curve and 175 K on the warming curve. The large thermal hysteresis has been reported by other authors[1,5] and may be an indication that charge ordering is a nucleation and growth process. Below the Néel temperature, the magnetisation is still higher than at room temperature and there are two possible causes of this effect: spin canting[6,7] or an inhomogeneous mixture of ferromagnetic and antiferromagnetic regions[2,4] within the same sample. We show that the latter is the correct explanation.

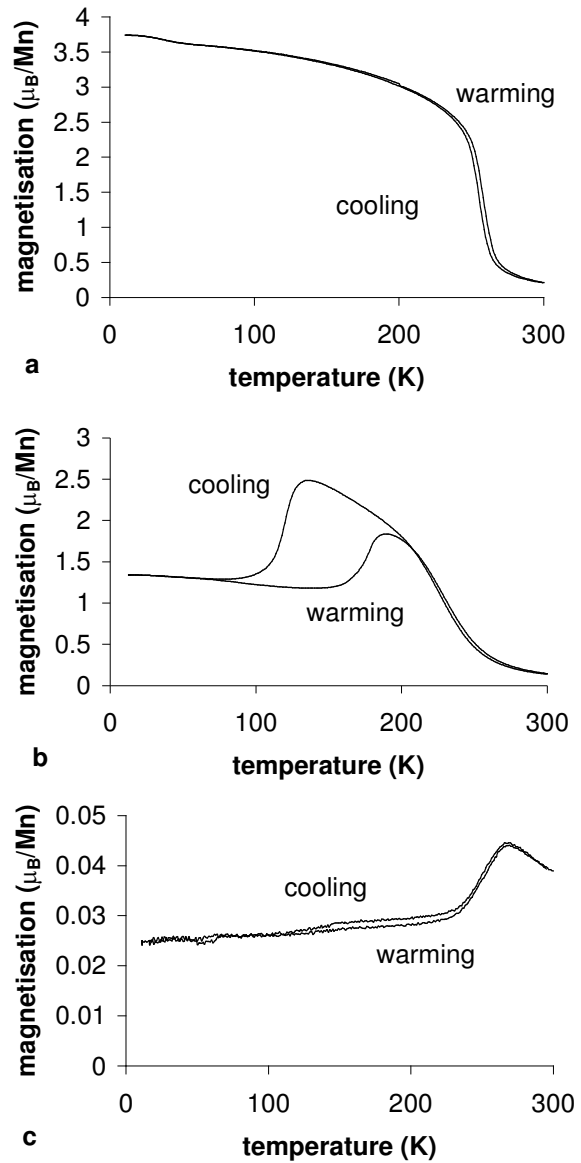

**Figure 1** Magnetisation against temperature curves for various compositions of polycrystalline $La_{1-x}Ca_xMnO_3$ under an applied field of 1 T. **a**, $x = 0.3$ shows a conventional paramagnetic to ferromagnetic phase transition (with a Curie temperature of 256 K measured at the inflexion point) and approaches the saturation moment of 3.7 $\mu_B$/Mn. The small hysteresis is almost certainly due to thermal inertia in the cryostat. **b**, $x = 0.5$ appears to show two transitions as the temperature is lowered: first a paramagnetic to ferromagnetic transition at 233 K and then an antiferromagnetic transition at 120K measured on the cooling curve and 175 K on the warming curve. **c**, $x = 0.66$ shows a paramagnetic to antiferromagnetic transition at 250 K.

According to the conventional model[8,9,10] for the low temperature phase transitions in manganites, the antiferromagnetic phase is associated with a spontaneous ordering of $Mn^{3+}$ and $Mn^{4+}$ ions. If the room temperature cell is indexed[11] as orthorhombic Pnma ($a$ = 5.42 Å, $b$ = 7.65 Å, $c$ = 5.44 Å), the lattice distortion caused by this charge ordering gives rise to extra reflections at positions $(h+q,k,l)$ at low temperatures[12] in a diffraction pattern where $q$ is the wavevector of the modulation. $q$ has been seen to vary with both composition and temperature[13] but in this paper, we consider only the commensurate phase $q$ = 1/2. If an objective aperture is placed over one of these reflections, parts of the sample that charge order appear bright in an image.

The $La_{0.5}Ca_{0.5}MnO_3$ sample was prepared for transmission electron microscopy by conventional mechanical polishing and argon ion thinning at liquid nitrogen temperatures. Large area (10 × 10 μm) energy dispersive x-ray analysis using a scanning electron microscope showed that the La/Ca ratio was constant to within the error of the measurements ($\delta x = \pm 2\%$). Transmission electron microscopy was undertaken using a 300 kV Philips CM300 microscope equipped with a field emission gun to provide a coherent source of electrons, a biprism used for holography and a 'Lorentz' lens (a strong objective mini lens) which allows the magnetic structure of a specimen to be observed in near-zero (less than 0.02 T) magnetic field. A liquid nitrogen specimen stage was used to cool the samples to 90 K.

Two methods were used to image the magnetic structure of the material: Fresnel imaging and electron holography. In the Fresnel method[14], the sample is imaged out of focus with a coherent beam of electrons. Lorentz forces in the sample cause magnetic domain walls to appear[15] as bright interference fringes where the electrons converge and dark regions where the electrons diverge.

In electron holography[16], an electron biprism (a positively charged wire) is inserted into the column of the microscope and used to interfere a reference wave passing through vacuum with one passing through the sample. A digital reconstruction allows both the amplitude and phase of the exit wavefunction to be determined directly.

Taking coordinates $x$ and $y$ normal to the beam direction $z$, the phase change on passing through the specimen is given by[16]

$$\phi(x,y) = C_E \int V_0(x,y,z)\,dz - \frac{2\pi e}{h} \iint \mathbf{B}_\perp(x,y,z) \cdot \mathbf{dS} \quad (1)$$

where $C_E = \frac{2\pi e}{\lambda}\left(\frac{eV+mc^2}{eV(eV+2mc^2)}\right) = 6.523 \times 10^6 \, m^{-1}V^{-1}$ for $V$ = 300 kV, $e$ is the electron charge, $m$ is the electron rest mass, $V$ is the acceleration voltage = 300 kV, $c$ is the speed of light, $\lambda$ is the wavelength of an electron = 0.0197 Å at 300 kV, $V_0(x,y,z)$ is the mean inner potential, $\mathbf{B}_\perp(x,y,z)$ is the component of magnetic flux density normal to the electron beam and $\mathbf{dS}$ is an element of vector area normal to the beam direction. This formula holds true provided that the specimen is not diffracting strongly. This is ensured by tilting away from any major zone axes.

By comparing holograms taken at room temperature (above the Curie temperature, $T_C$) with those taken at 90 K (below $T_C$), we can rearrange equation (1) to give a measure of the absolute value of the magnetisation in a way that does not depend on the unknown thickness of the sample:

$$\mu_0 \mathbf{M}_\perp(x,y) = \mathbf{B}_\perp(x,y) = -\frac{h}{2\pi e} C_E V_0 \frac{1}{\phi_{T>T_C}(x,y)}\begin{pmatrix}\partial/\partial y \\ -\partial/\partial x\end{pmatrix}\{\phi_{T<T_C}(x,y) - \phi_{T>T_C}(x,y)\} \quad (2)$$

This formula assumes that: (i) $V_0$ is constant throughout the sample and has the value of 23.73 V calculated from electron scattering factors[17], (ii) $\mathbf{B}_\perp(x,y)$ is constant throughout the thickness of the sample and (iii) the stray field is negligible in comparison with the flux density within the sample.

In Fig. 2 we show ferromagnetism and charge order coexisting. The field of view of a hologram is limited by the size of the interference region created by the biprisim and so we present Fig. 2(a) as a montage of regions investigated by holography (colour) overlaid on a Fresnel image. This image was taken using the Lorentz lens to focus the electrons. Typically, this lens places the specimen in a vertical magnetic field of less than 0.02T. The Fresnel image shows long straight black and white lines in the lower left hand corner. These are magnetic domain walls which match the coloured domains derived from holography. (The black curves at the top right are bend contours.) Three grains are present in 2(a) labeled 1, 2 and 3. In grain 1, the orthorhombic **b** axis lies approximately in the plane and 180° domains dominate whereas in grain 2, the **b** axis lies approximately normal to the film and there is a mixture of 90° and 180° domains. This change in magnetic behaviour probably arises because of the change in the orientation of the unit cell and the shape anisotropy caused by the thin specimen[18].

The magnetisation measured over 100 × 100 nm squares at several positions in several domains in grains 1 and 2 was 3.4 $\mu_B$/Mn on average with a standard deviation of 0.5 $\mu_B$/Mn. If it is assumed that this variability comes only from random errors, we can give the average local magnetisation as 3.4 ± 0.2 $\mu_B$/Mn. The holograms showed that the stray field in the vacuum near the specimen edge was approximately 5% of the flux density in the sample so the stray field may be safely neglected. The mean inner potential $V_0$ was derived from electron scattering factors assuming that all the atoms are neutral and no bonding takes place. This assumption is likely to overestimate the magnetisation possibly by as much as 10%. In comparing this magnetisation with the saturation moment, we assume that the magnetisation lies in the plane of the specimen as would be expected by considering the shape anisotropy. The magnetisation of $3.4 \pm 0.2\,\mu_B$/Mn measured here agrees well with the value of $3.5\,\mu_B$/Mn that would be expected if the sample were fully ferromagnetic and is significantly larger than the bulk value of $1.25\,\mu_B$/Mn measured by magnetometry.

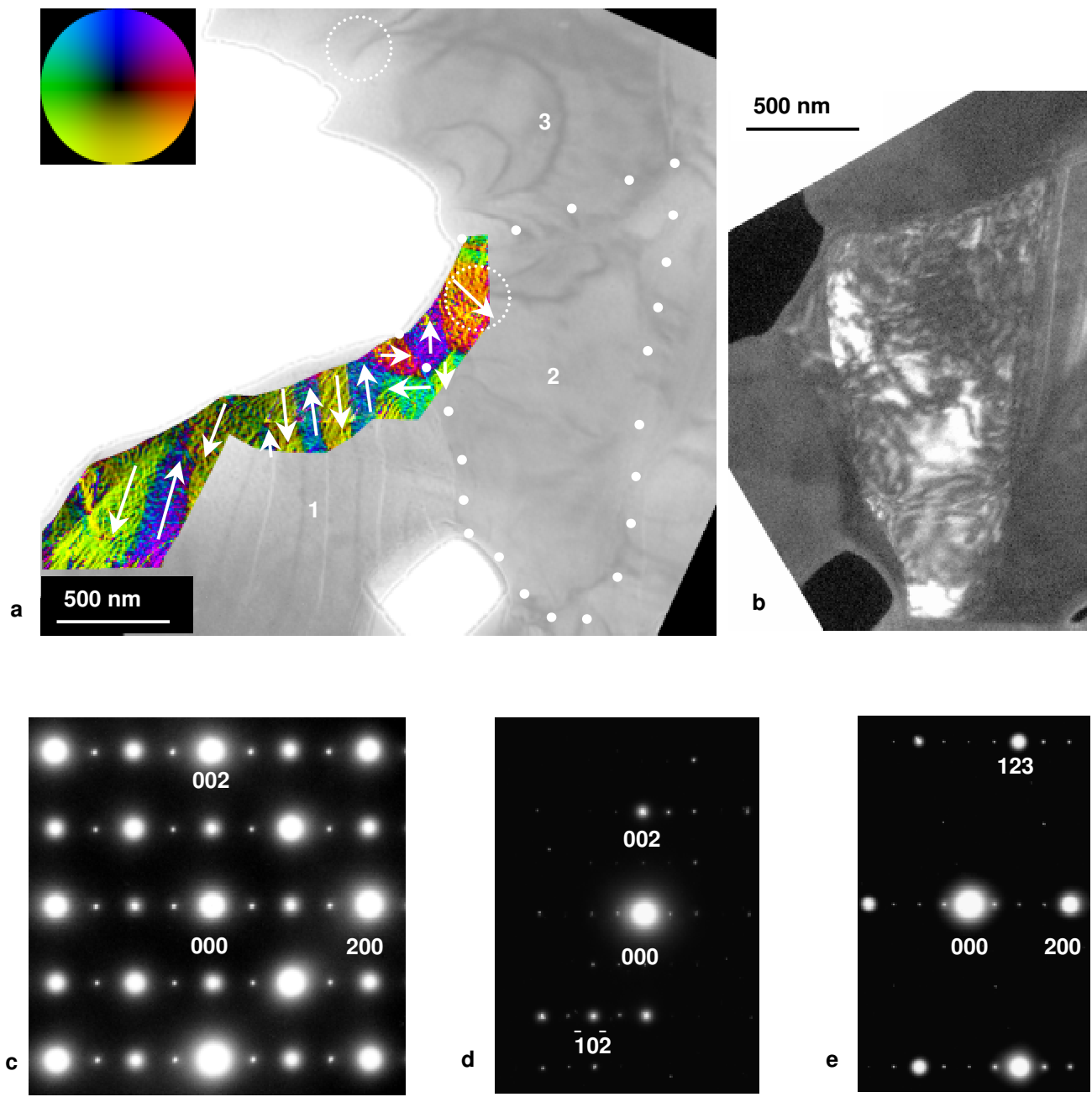

**Figure 2** Coexistence of ferromagnetism and charge order. Three grains are present in the image and are labelled 1, 2 and 3. The grain boundary is indicated by the white dots. **a** A colour image of the magnetic structure derived from available holography data overlaid on a Fresnel image. The colour wheel indicates the direction of the local magnetisation (marked with arrows). The apparent texture within each domain is likely to be an artefact of the Fourier filtering process used to reconstruct the phase of the exit wavefunction. **b** A dark field image on the same scale from grain 2 taken using the 1/2,0,2 charge order reflection. **c** A diffraction pattern taken from the circled region in 2 looking down the [010] axis with charge ordered reflections at $h+1/2,k,l$. **d** A diffraction pattern from the same region as **c** tilted away from the [010] zone axis by a few degrees. The charge ordered reflections persist but reflections of type 00l (for l odd) are now absent as they are kinematically forbidden by the Pnma space group. **e** A diffraction pattern from the circled region in grain 3 looking down the [03$\bar{2}$] zone axis. Charge ordered reflections can be seen at $h+1/2,k,l$. Holography showed that this region had no net magnetisation. The very weak reflections between the main rows are from a twin variant of this axis, the [214] zone axis.

Holography showed that grain 3 had no net magnetisation as did several other grains in other regions of the specimen. Fig. 2(e) shows a diffraction pattern taken from the circled region in grain 3 showing charge ordering. This is the expected charge ordered, non-ferromagnetic phase. We conclude that at 90 K, the sample is composed of an inhomogeneous mixture of ferromagnetic regions and regions with no net magnetisation. Given that the bulk magnetisation at this temperature is 1.25 $\mu_B$/Mn compared with the theoretical value of 3.5 $\mu_B$/Mn, around 36% of the sample is ferromagnetic.

Fig. 2(b) shows a dark field image from grain 2 taken using the (1/2,0,2) (charge ordered) reflection and Fig. 2(c) shows the associated diffraction pattern taken from the region circled in grain 2 showing charge ordered reflections at positions ($h+1/2,k,l$). These images were taken with the Lorentz lens turned off and the electrons focussed with the main objective lens which places the sample in a vertical magnetic field of around 2T. This will tend to cause the magnetisation to rotate towards the vertical and possibly cause the charge ordered regions to shrink. Despite this, both the diffraction pattern and the dark field image show charge order present in the region of the sample where holography revealed ferromagnetic domains. The simplest explanation for the dark field contrast seen in grain 2 (Fig. 2(b)) is that the bright regions are charge ordered and ferromagnetic and the dark regions are ferromagnetic and charge disordered.

We can rule out the possibility that the ferromagnetic charge ordered phase is in fact charge ordered antiferromagnetic regions coexisting with ferromagnetic metallic regions on a very small scale. The spatial resolution of the dark field image is determined by the size of the objective aperture and the specimen drift rate and is about 2 nm. The spatial resolution of the holograms is determined by the size of the window used in the Fourier filtering process as part of the reconstruction process and is about 20 nm. The large white patch in Fig. 2(b) which overlaps the magnetic region in Fig 2(a) is about 100 nm in size. Thus if a small scale phase separation takes place, the coexisting phases must be less than 2 nm or around 4 unit cells in size and such a scenario must be regarded as mesoscopic texture rather than the separation of two thermodynamically stable phases as argued in ref. 3. Furthermore, if this were the case, one would expect the magnetisation in the overlapping region to be approximately halved. In fact, the magnetisation in this region is $3.5 \pm 0.2$ $\mu_B$/Mn in excellent agreement with the spin aligned value of 3.5 $\mu_B$/Mn. We can similarly rule out the possibility that this region may be a ferromagnetic phase on top of a separate non-ferromagnetic, charge ordered phase since the charge ordered phase would have to be less than 2 nm or around 2 unit cells thick to produce a magnetisation within the error limits quoted. It is highly unlikely that charge order has developed over such a small length scale because part of the region of overlap is very bright in the dark field image (Fig. 2(b)) indicating that charge order is well established here. We have also taken a number Fresnel images and diffraction patterns from other areas of the sample and have seen magnetic domain walls and charge ordering in the same region which is further evidence for this unexpected phase.

In tilting the sample, we found that charge ordered reflections perpendicular to a systematic row were far stronger for those parallel to the systematic row. Fig. 2(d) shows a systematic row from the same region as Fig. 2(c) with strong charge order reflections perpendicular to the row. This is strong evidence that charge order is predominantly a transverse modulation of the atoms as discussed by Radaelli[11] based on evidence from neutron powder diffraction.

Using electron microscopy, we have shown that at 90 K, $La_{0.5}Ca_{0.5}MnO_3$ forms an inhomogeneous mixture of ferromagnetic and zero-moment (presumably antiferromagnetic[1]) regions. Each region extends for several microns and can span several crystallographic grains. We have also shown that charge order occurs not only in regions with no net magnetisation, but can also occur in ferromagnetic regions. This charge ordered, ferromagnetic phase is not predicted by conventional models[8,9,10]. We suggest that the band gap which opens up below the charge ordering transition temperature can be small enough that the valence electrons still have enough mobility to promote ferromagnetism via double exchange and yet large enough to produce a charge density wave. This seems reasonable as ferromagnetic coupling in the manganites should only require nearest neighbour hopping. Furthermore, weak charge ordering may even be possible deep within the ferromagnetic regime since Zuo & Tao[19] have observed diffuse charge order reflections in $La_{2/3}Ca_{1/3}MnO_3$. We further suggest that the band gap is locally modified by strain within each grain and this results in the rich variety of phases that can coexist in the manganites. This may also explain the strong dependence of $La_{0.5}Ca_{0.5}MnO_3$ properties on grain size[5]. Therefore manganite phase diagrams based on composition alone paint an inadequate picture of manganite physics and miss the richness and complexity that we have demonstrated here.

**Acknowledgements**
We should like to thank R. Dunin-Borkowski for his help with electron holography, D. Vowles for his help in using the scanning electron microscope, A. Williams and J. P. Attfield for helpful discussions and A. J. Millis and P. B. Littlewood for information on the nature of charge order. This work was funded by the EPSRC, UK and by the Royal Society.

Correspondence should be addressed to J.C.L. (e-mail: james.loudon@physics.org).